\documentclass[12pt,preprint]{aastex}

\begin{document}
\newcommand{\calu}{{\cal U}}
\newcommand{\calq}{{\cal Q}}
\newcommand{\bx}{{\rm \bf x}}
\newcommand{\bk}{{\bar{\kappa}}}
\title{Constraints on holographic dark energy models using the
differential ages of passively evolving galaxies}
\author{Ze-Long Yi, Tong-Jie Zhang}
\affil{Department of Astronomy, Beijing Normal University, Beijing,
100875, P.R.China; tjzhang@bnu.edu.cn}

\begin{abstract}

Using the absolute ages of passively evolving galaxies observed at
different redshifts, one can obtain the differential ages, the
derivative of redshift $z$ with respect to the cosmic time $t$ (i.e.
${\rm d} z/{\rm d}t$). Thus, the Hubble parameter $H(z)$ can be
measured through the relation $H(z)=-({\rm d} z/{\rm d}t)/(1+z)$. By
comparing the measured Hubble parameter at different redshifts with
the theoretical one containing free cosmological parameters, one can
constrain current cosmological models.
In this paper, we use this method to present the constraint on a
spatially flat Friedmann-Robert-Walker Universe with a matter
component and a holographic dark energy component, in which the
parameter $c$ plays a significant role in this dark energy model.
Firstly we consider three fixed values of $c$=0.6, 1.0 and 1.4 in
the fitting of data. If we set $c$ free, the best fitting values are
$c=0.26$, $\Omega _{\rm m0}=0.16$, $h=0.9998$. It is shown that the
holographic dark energy behaves like a quintom-type at the $1\sigma$
level. This result is consistent with some other independent
cosmological constrains, which imply that $c<1.0$ is favored. We
also test the results derived from the differential ages using
another independent method based on the lookback time to galaxy
clusters and the age of the universe. It shows that our results are
reliable.

\end{abstract}

{\bf Keywords}: Cosmology; cosmological parameters; holographic dark energy;
the differential ages of galaxies

{\bf PACS numbers}: 98.80.-k, 98.80.Es, 98.80.Cq, 95.35.+d\\

\section{Introduction}
The type Ia Supernova (SN Ia) observations indicate that the
expansion of our Universe is accelerating today [1,2], which can be
explained by an exotic component, referred to as dark energy, with a
large negative pressure.
The most obvious candidate of dark energy is the cosmological
constant [3]. In addition, many other cosmological models such as
the quintessence [4], the Chaplygin Gas [5], the braneworld models
[6], the Cardassian models [7] are proposed to explain the
accelerating expansion of the Universe.
Recently, Cohen et al. [8] suggested the holographic dark energy
model to reconcile the breakdown in quantum field theory caused by
the conflict between the effective field theory and the so-called
Bekenstein entropy bound [9,10]. For an effective field theory, in a
box of size $L$ with UV cut-off $\Lambda$, the entropy $S$ scales
extensively, $S\sim L^3\Lambda ^3$. But for the Bekenstein entropy
bound, $S$ is non-extensive, $S\leq S_{\rm BH}\equiv L^2/8G$. This
is postulated from the peculiar thermodynamics of a black hole.
Cohen et al. also pointed out that, if the zero-point energy density
$\rho _{\rm H}$ is related to a UV cut-off, the total energy of the
whole system with size $L$ should not exceed a black hole of the
same size, i.e. $\rho _{\rm H}\leq L^{-2}/8\pi G$. It means that the
maximum entropy is in order of $S_{\rm BH}^{3/4}$. A short distance
UV cut-off and a long distance IR cut-off have been connected
together. If we apply this to the Universe, the vacuum energy
related to the holographic principle [11] is called the holographic
dark energy. Thus the holographic dark energy density $\rho _{\rm H}$
is written as
\begin{equation}
\rho _{\rm H}=\frac{3c^2}{8\pi G}L^{-2},\label{eq1}
\end{equation}
where $c$ is a numerical constant. As proposed by Li [12], we take
the scale $L$ as the size of the future event horizon
\begin{equation}
R_{\rm h}(a)=a\int_t^{\infty}\frac{{\rm
d}t}{a(t)}=a\int_a^{\infty}\frac{{\rm d}a'}{Ha'^{2}},\label{eq2}
\end{equation}
where $a$ is the scale factor of the Universe, $t$ is the cosmic
time and $H$ is the Hubble parameter. Such a dark energy model has
been constrained through many kinds of astronomical observations,
including
SN Ia [13], X-ray gas mass fraction of galaxy clusters [14] and a
joint analysis of SN Ia and CMB and LSS observational data [15].

It is noted that $c$ is one key parameter in the holographic dark
energy model [13-15], which determines the evolutionary behavior of
the space-time and the ultimate fate of the Universe.
Huang et al. [13] suggested $c=0.21$ through a direct fit to the
present available SN Ia data. Zhang et al. [15] showed $c=0.81$ by
combining SN Ia and CMB and LSS. Kao et al. [16] showed that the
reasonable result is $c\sim 0.7$ by calculating the average equation
of state and the angular scale of the oscillation from the BOOMERANG
and WMAP data on the CMB observation.

Unlike the observational data mentioned above, Simon et al. [17]
used the relative ages of a sample of passively evolving galaxies
and supernova data to reconstruct the potential of dark energy. They
showed that the reconstructed potentials from galaxy ages and SN are
consistent. In this paper, we attempt to apply the differential ages
of galaxies [17,18] to the constraint on the holographic dark energy
model.
If we take some fixed values $c$=0.6, 1.0 and 1.4, the corresponding
results of $h$ and $\Omega _{\rm m0}$ are not too far from each
other. If we set all the parameters free, the best fitting result is
$\Omega _{\rm m0}=0.16$, $h=0.9998$, $c=0.26$, corresponding to a
quintom-type dark energy at the $1\sigma$ level. This result is
consistent with some other independent cosmological tests. In
addition, we attempt to test the fitting results using another
independent method based on the data for the lookback time to galaxy
clusters [19] and the age of the universe. It is shown that our
method is acceptable.

This paper is organized as follows. In Sec.2, we briefly review the
holographic dark energy model. Differential ages of galaxies and
constraints on holographic dark energy model are presented in Sec.3,
while another independent method based on the lookback time to
galaxy clusters and the age of the universe is used to test our
conclusions in Sec.4. The deceleration parameter $q$ and equation of
state $\omega$ are discussed in in Sec.5. Conclusions and
discussions are given in the last section.

\section{The Holographic Dark Energy Model}
Recent observations indicate that our Universe has a flat geometry
[20]. Here we consider a spatially flat Friedmann-Robert-Walker
Universe with a matter component $\rho_{\rm m}$ (including both
baryon mater and cold dark matter) and  a holographic dark energy
component $\rho_{\rm H}$, which means $\Omega_{\rm m}+\Omega_{\rm
H}=1$. Therefore, the Friedman equation reads
\begin{equation}
 \frac{3H^2}{8\pi G}=\rho_{\rm m}+\rho_{\rm H},\label{eq3}
\end{equation}
or taking this form equivalently
\begin{equation}
\frac{H^2}{H_{\rm 0}^2}=\Omega _{\rm m0}a^{-3}+\Omega _{\rm
H}\frac{H^2}{H_{\rm 0}^2},\label{eq4}
\end{equation}
where $\Omega _{\rm m0}=\rho_{\rm m0}/\rho_{\rm c}$, $\Omega_{\rm
H}=\rho_{\rm H}/\rho _{\rm c}$, $\rho_{\rm c}=3H_{\rm 0}^2/8\pi G$
and $H_{\rm 0}=100h$ km s$^{-1}$ Mpc$^{-1}$ is the Hubble constant.
Combining Eq.(\ref{eq1}) and Eq.(\ref{eq2}), we derive
\begin{equation}
\int_a^\infty \frac{d{\rm ln}a'}{Ha'}=\frac{c}{Ha\sqrt{\Omega_{\rm
H}}}.\label{eq5}
\end{equation}
Meanwhile we notice that Eq.(\ref{eq4}) can be deduced to
\begin{equation}
\frac{1}{Ha}=\sqrt{a(1-\Omega_{\rm H})}\frac{1}{H_{\rm
0}\sqrt{\Omega_{\rm m0}}}.\label{eq6}
\end{equation}
Substituting Eq.(\ref{eq6}) into Eq.(\ref{eq5}) and taking
derivative with respect to ln$a$ on both sides, we get

\begin{equation}
\frac{d\Omega_{\rm H}}{d{\rm ln}a}=\Omega_{\rm H}(1-\Omega_{\rm
H})(1+\frac{2}{c}\sqrt{\Omega_{\rm H}}), \label{eq7}
\end{equation}
which describes the dark energy's evolution along with the
factor ln$a$ completely. And it can also be solved analytically as follows
\begin{equation}
{\rm ln}\Omega_{\rm H}-\frac{c}{2+c}{\rm ln}(1-\sqrt{\Omega_{\rm
H}})+\frac{c}{2-c}{\rm ln}(1+\sqrt{\Omega_{\rm
H}})-\frac{8}{4-c^2}{\rm ln}(c+2\sqrt{\Omega_{\rm H}})=-{\rm
ln}(1+z)+y_0,\label{eq8}
\end{equation}
where $y_0$ can be determined by given $\Omega _{\rm H0}$ and $c$. From
Eq.(\ref{eq4}), it is easy to get the theoretical Hubble parameter
\begin{equation}
H(z)=H_0\sqrt{\frac{\Omega_{\rm m0}(1+z)^3}{1-\Omega_{\rm
H}}},\label{eq9}
\end{equation}
where $\Omega _{\rm H}$ is a function of $\Omega _{\rm H0}$ and $c$, i.e.,
$\Omega _{\rm H}=f(\Omega _{\rm H0},c)$ and $\Omega _{\rm m0}+\Omega _{\rm H0}=1$.

\section{Differential Ages of Galaxies and Constraints on Holographic Dark Energy}

The Hubble parameter depends on the differential age of the Universe
as a function of $z$ in this form
\begin{equation}
H(z)=-\frac{1}{1+z}\frac{{\rm d}z}{{\rm d}t},\label{eq10}
\end{equation}
so $H(z)$ can be directly measured through a determination of ${\rm
d}z/{\rm d}t$.  By comparing the measured Hubble parameter at
different redshifts with the theoretical one containing some free
cosmological parameters, we can constrain current cosmological
models.

In this paper, we adopt data from [17] where the new publicly
released Gemini Deep Survey (GDDS) survey [21] and archival data
[22-26] were used to infer $dz/dt$ and the shape as well as redshift
dependence of dark energy potential were constrained. The data
contain the absolute ages of 32 passively evolving galaxies, which
are further divided into three subsamples. The first subsample was
composed of 10 field early-type galaxies, after discarding galaxies
for which the spectral fit indicated an extended star formation. The
ages of this sample were derived using the SPEED models [27]. The
second subsample was composed of 20 old passive galaxies from GDDS.
The GDDS old sample was reanalyzed using SPEED models and the ages
of this sample were obtained within 0.1Gyr of the GDDS collaboration
estimate. The third subsample consisted of two reddest radio
galaxies 53W091 and 53W069 in the survey of Windhorst et al.
[24-26].
Then 8 numerical values of differential ages $dz/dt$, equivalently
$H(z)$, were obtained. The detailed procedure on calculating the
differential ages can be found in [17].

\begin{figure}
\centering \epsscale{0.9}
\includegraphics[width=15.5cm]{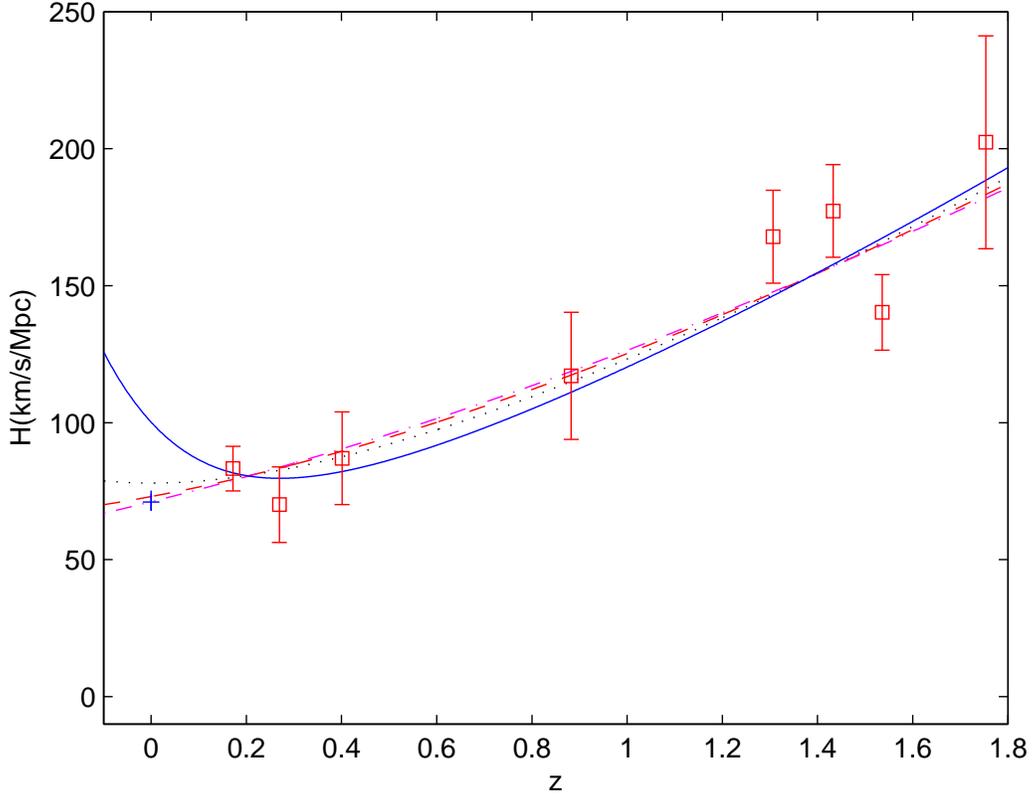}
\caption{$H$ as a function of $z$ for the holographic dark energy
models. The dot line corresponds to $c=0.6$, $\Omega_{\rm m0}=0.24$,
$h=0.78$, the dash $c=1.0$, $\Omega_{\rm m0}=0.23$, $h=0.74$ and the
dash-dot $c=1.4$, $\Omega_{\rm m0}=0.21$, $h=0.71$. The solid line
represents the case of considering $c$ free, $c=0.26$, $\Omega_{\rm
m0}=0.16$ and $h=0.9998$. The open squares with $1\sigma$ error bar
are the derived values from the differential ages of galaxies, while
the plus symbol stands for the result of WMAP
$h=0.71_{-0.03}^{+0.04}$ [29].} \label{fig1}
\end{figure}

The parameters of the holographic dark energy model can be
determined by minimizing
\begin{equation}
\chi^2=\sum_i\frac{[H_{\rm th}(z_{i})-H_{\rm de}(z_{i})]^2}{\sigma
_i^2},\label{eq11}
\end{equation}
where $H_{\rm th}(z_{i})$ is the theoretical Hubble parameter as
Eq.(\ref{eq9}), $H_{\rm de}(z_{i})$ is the Hubble parameter derived
from the differential ages as Eq.(\ref{eq10}) and $\sigma _i$ is the
corresponding $1\sigma$ error bar.
All the fitting results are listed in Table.1 for the fixed choice
of $c$=0.6, 1.0 and 1.4. We can see that the result of $c=0.6$ is
roughly consistent with Chang et al: $c=0.61$, $\Omega_{\rm
m0}=0.24$, $h=0.73$ by X-ray gas mass fraction of galaxy clusters
[14]. For the case of $c=1.0$, Huang \& Gong [13] suggested $\Omega
_{\rm m0}=0.25_{-0.03}^{+0.04}$ using 157 gold sample in [28] and
$\Omega _{\rm m0}=0.25\pm 0.03$ using 186 gold and silver SN sample.
If $c$ is set to be free, the best fitting result is $c=0.26\pm
0.37$, $\Omega_{\rm m0}=0.16\pm 0.20$, $h=0.9998\pm 0.6903$ and
$\chi _{\rm min}^2=7.61$. The best fitting value of $\Omega_{\rm
m0}$ is quite smaller and $h$ is quite larger than the result of
WMAP [29]. The fitting value of $c$ is roughly consistent with
$c=0.21$ in [13] although the value of $\Omega_{\rm m0}$ is smaller
than $\Omega_{\rm m0}=0.46$ there. But Zhang et al. [15] showed
$c=0.81$, $\Omega_{\rm m0}=0.28$, $h=0.65$ by combining SN Ia data
and the CMB and LSS data.
We plot $H$ as a function of $z$ in Fig.\ref{fig1} using the best
fitting values.

\begin{deluxetable}{cccc}
\tablecolumns{4} \tablewidth{0pc} \tablecaption{The
fitting results for the fixed cases: $c$=0.6, 1.0 and 1.4.}
 \tablehead{ &
\colhead{$c=0.6$} & \colhead{$c=1.0$} & \colhead{$c=1.4$}}
\startdata
  $\Omega_{\rm m0}$ & $0.24\pm 0.08$ & $0.23\pm 0.08$ & $0.21\pm 0.09$ \\
  $h$                & $0.78\pm 0.09$ & $0.74\pm 0.08$ & $0.71\pm 0.07$ \\
  $\chi_{\rm min}^2$    &     $7.62$     &     $7.61$     &     $7.59$
 \label{table1}
\enddata
\end{deluxetable}

\begin{figure}
\centering \epsscale{0.9}
\includegraphics[width=15.5cm,height=16.0cm]{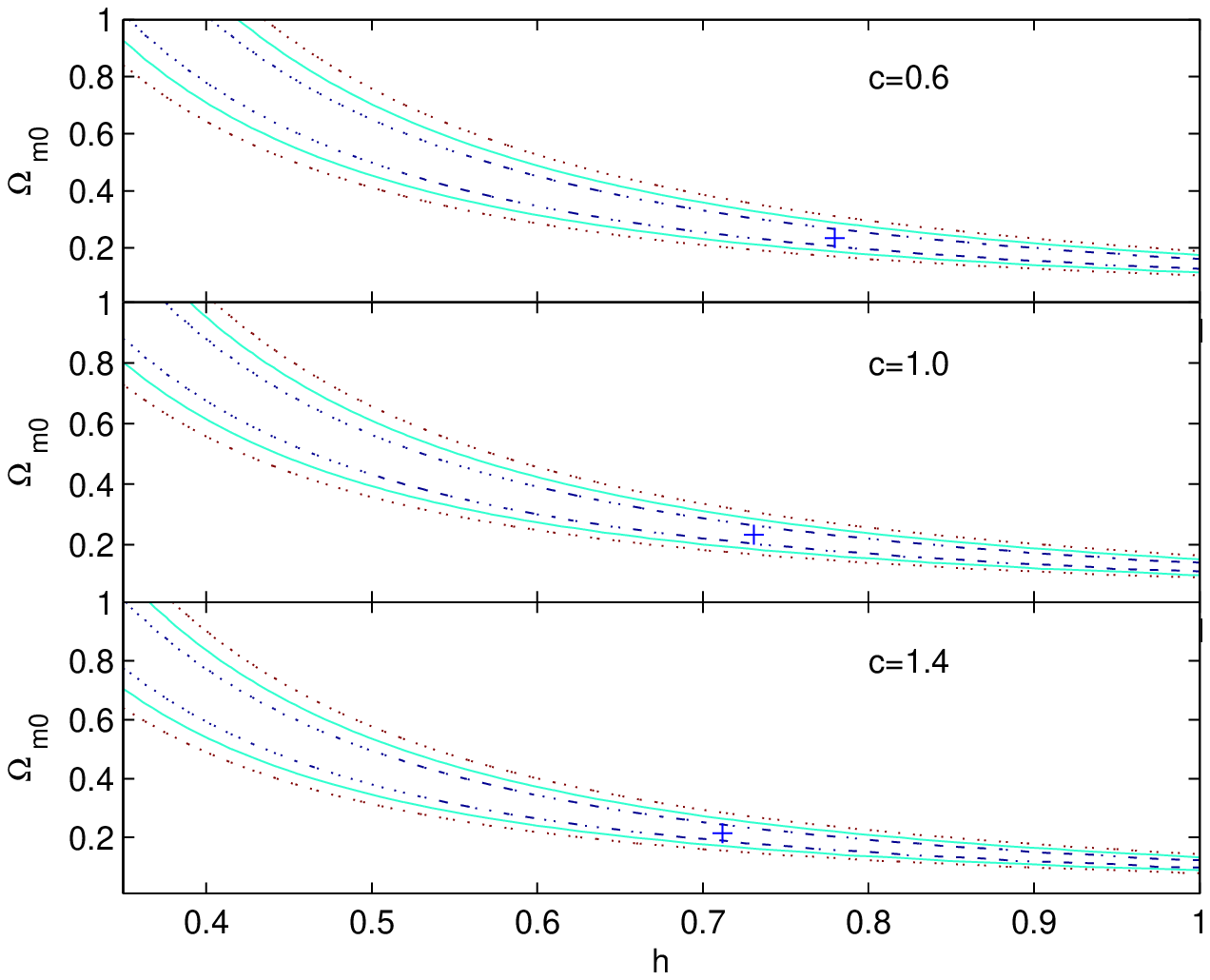}
\caption{Confidence regions in the $\Omega_{\rm m0}-h$ plane for
$c=$0.6, 1.0 and 1.4. The dash-dot line corresponds to the
confidence level of 68.3\%, the solid 95.4\% and the dash 99.7\%
respectively. The cross stands for the respective best fitting
values.} \label{fig2}
\end{figure}

\begin{figure}
\centering \epsscale{0.9}
\includegraphics[width=15.5cm]{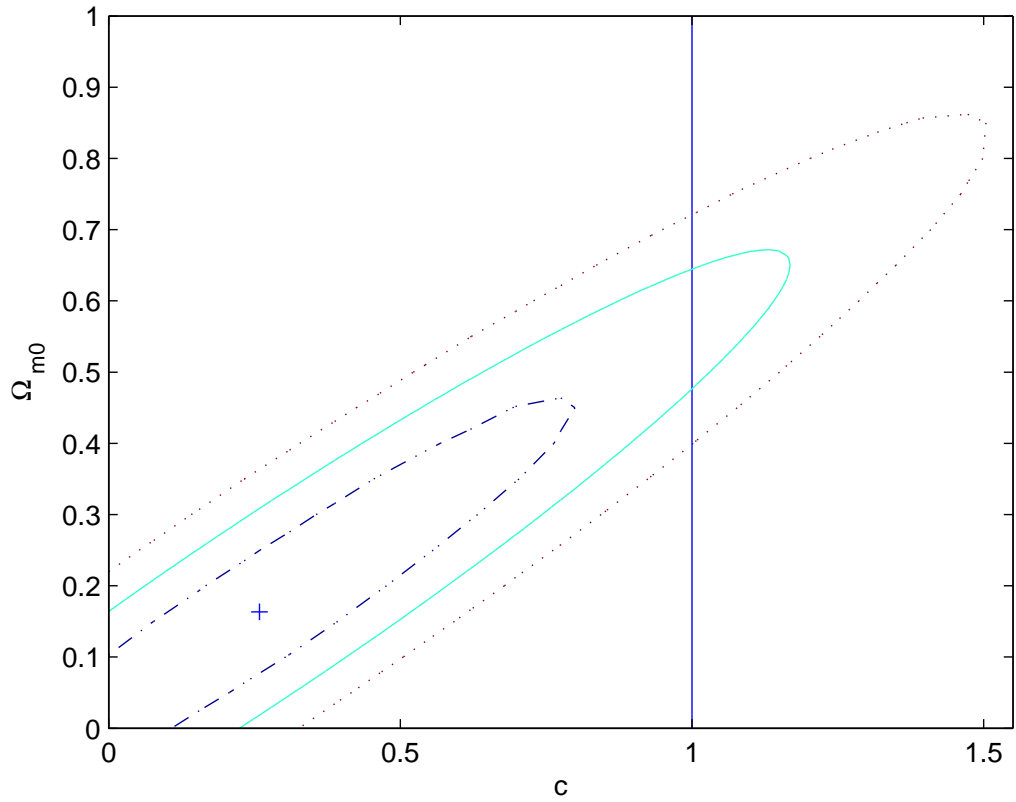}
\caption{Same as Fig.\ref{fig2}, but in the $\Omega_{\rm m0}-c$
plane marginalizing the likelihood function over $h$ and setting $c$
free.} \label{fig3}
\end{figure}

The confidence regions in the $\Omega_{\rm m0}-h$ plane are plotted
in Fig.\ref{fig2} for $c$=0.6, 1.0 and 1.4 respectively.
We can see that $\Omega_{\rm m0}$ and $h$ are constrained a little
poorly. The confidence regions in the $\Omega_{\rm m0}-c$ plane for
setting $c$ free, after marginalizing the likelihood function over
$h$, are plotted in Fig.\ref{fig3}. And what we have to point out is
that we've assumed an approximate Gaussian prior for a convenience
during this work.
The contour plots show $c\lesssim 0.81$ at the $1\sigma$ level
although we can not exclude the existence of $c>1.0$ at $95.4\%$
confidence level.
This is in good agreement with some other independent cosmological
constraints [13,14].

\section{An Independent Constraints Using the Lookback Time to Galaxy Clusters and
the Age of the Universe}


In order to guarantee the reliability of the method above, we make a
further independent constraint on the model using the lookback time
to galaxy clusters [19] and the age of the universe [30]. Dalal et
al. [31] proposed the use of the lookback time to high redshift
objects to constrain cosmological models. Capozziello et al. [19]
constrained the quintessence model, the parametric density model and
the curvature quintessence model using the lookback time to galaxy
clusters and the age of the universe.

\begin{figure}
\centering \epsscale{0.9}
\includegraphics[width=15.5cm]{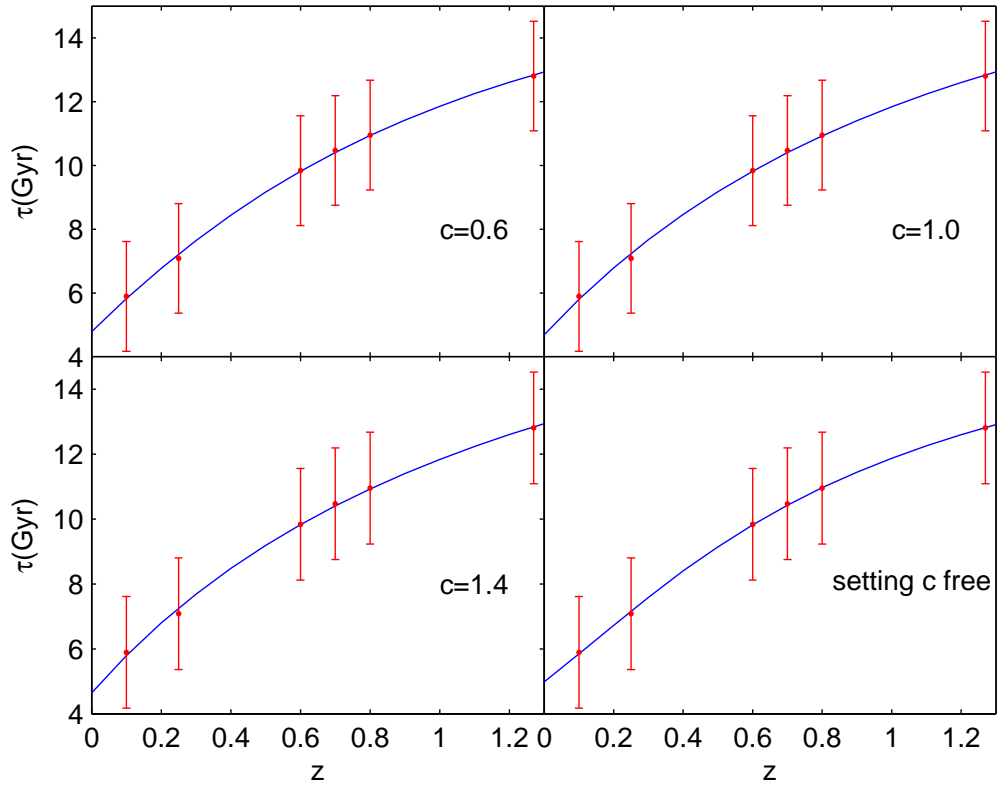}
\caption{$\tau$ as a function of $z$ using the best fitting results.
The data points with $1\sigma$ error bar are observational values.}
\label{fig4}
\end{figure}

Following the reference [19], the lookback time $t_{\rm L}$ can be
defined as the difference between the age of the universe at present
and the age at redshift $z$
\begin{equation}
t_{\rm L}(z; \textbf{P})=\int_0^z\frac{dz'}{(1+z')H(z';
\textbf{P})},\label{eq12}
\end{equation}
where $\textbf{P}$ is a set of parameters characterizing a given
cosmological model and $H(z';\textbf{P})$ is the corresponding
theoretical Hubble parameter. For the holographic dark energy model,
$\textbf{P}=(H_0, \Omega_{\rm m0}=1-\Omega_{\rm H0}, c)$ and $H(z;
\textbf{P})=H_0\sqrt{\Omega_{\rm m0}(1+z)^3/(1-\Omega_{\rm
H}(\Omega_{\rm H0}, c))}$ which is identical to Eq.(\ref{eq9}).
The age of an object at redshift $z$ can be expressed as
\begin{equation}
t(z)=\int_z^\infty\frac{dz'}{(1+z')H(z';\textbf{P})}-\int_{z_{\rm
F}}^{\infty}\frac{dz'}{(1+z')H(z';\textbf{P})},\label{eq13}
\end{equation}
where $z_{\rm F}$ is the redshift at which the galaxy is assumed to
be formed. Using the definition of the lookback time in
Eq.(\ref{eq12}), we can express $t(z)$ in terms of $t_{\rm
L}(z_{\rm F})$ and $t_{\rm L}(z)$
\begin{equation}
t(z)=t_{\rm L}(z_{\rm F})-t_{\rm L}(z).\label{eq14}
\end{equation}
Thus we can estimate the observational lookback time $t_{\rm
L}^{obs}(z)$ as
\begin{equation}t_{\rm L}^{obs}(z)=t_{\rm L}^{obs}(z_{\rm
F})-t^{obs}(z)=(t_0^{obs}-t^{obs}(z))-(t_0^{obs}-t_{\rm
L}^{obs}(z_{\rm F}))=t_0^{obs}-t^{obs}(z)-df,\label{eq15}
\end{equation}
where $t_0^{obs}$ is the present estimated age of the universe,
$t^{obs}(z)$ is the age of one object at redshift $z$ and
$df=t_0^{obs}-t_{\rm L}^{obs}(z_{\rm F})$ is a defined \emph{delay
factor}.
Thus we could estimate values of $\tau^{obs}(z)=t_{\rm
L}^{obs}(z)+df=t_0^{obs}-t^{obs}(z)$ at redshift $z$ through an
estimate of $t_0^{obs}$ and $t^{obs}(z)$. The data of the estimated
cluster age $\tau^{obs}(z)$ at different redshift are adapted from
[19].
We compare the theoretically predicted and observed values of $\tau$
by minimizing $\chi^2$ defined in Eq.(\ref{eq11}).
The best fitting results are: $h=0.91,\Omega_{\rm m0}=0.14, df=4.79,
\chi_{\rm min}^2=0.0097$ for $c=0.6$; $h=0.83, \Omega_{\rm m0}=0.12, df=4.69, \chi_{\rm
min}^2=0.014$ for $c=1.0$; $h=0.80, \Omega_{\rm m0}=0.085, df=4.65,
\chi_{\rm min}^2=0.016$ for $c=1.4$; $h=1.17, \Omega_{\rm
m0}=0.11, c=0.30, df= 4.99, \chi_{\rm min}^2=0.0034$ for setting $c$ free.
We plot $\tau$ as a function of $z$ in
Fig.\ref{fig4} using the best fitting values.

We further find that the best fitting results for larger values of
$c=1.0$ and $1.4$ are not consistent with the confidence regions
derived from the differential ages of passively evolving galaxies,
while that of case $c=0.6$ is within the $2\sigma$ confidence
region. For the case setting $c$ free, the best fitting result from
the lookback time method is in good agreement with the that from
differential ages within 68.3\% confidence region. Similar to the
results from differential ages, the best fitting value of $h$ is
also larger than the result of WMAP [29]. In addition, the values of
$\chi_{\rm min}^2$ also reveals that there exist better fitting
results for smaller $c$. Therefore, it seems that a smaller $c$ is
more suitable.

\begin{figure}
\centering \epsscale{0.9}
\includegraphics[width=15.5cm]{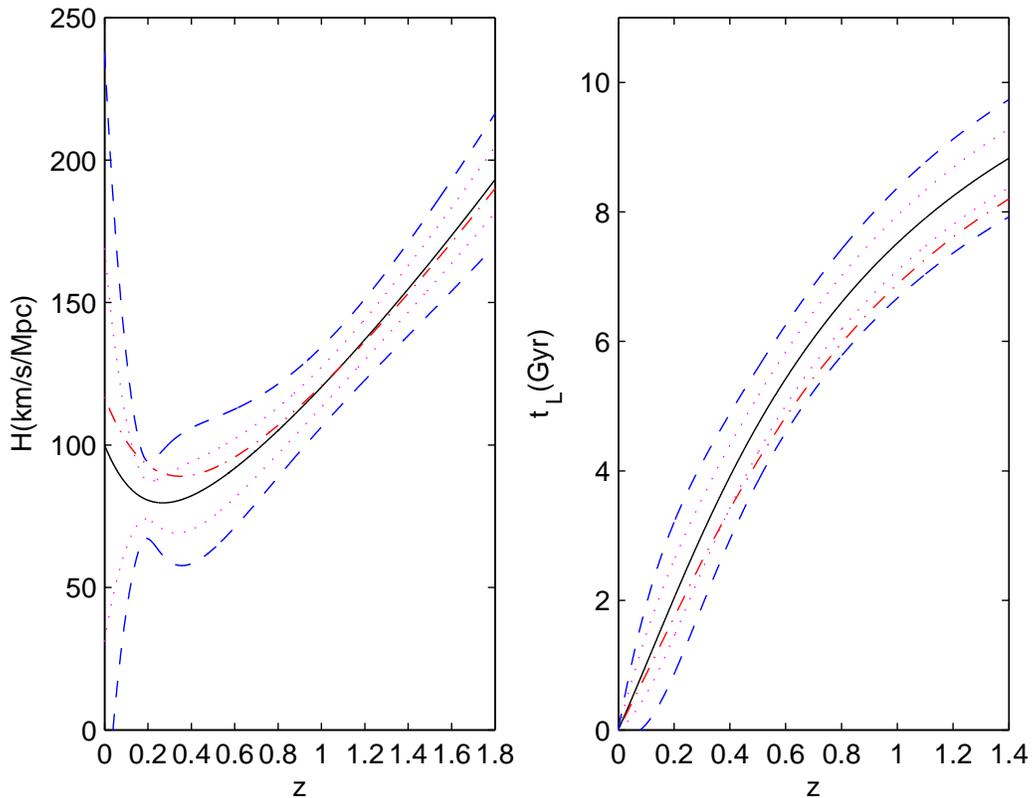}
\caption{ $H$(the left panel) and $t_{\rm L}$ (the right panel) as a
function of redshift $z$. The solid curves stand for $H$ for best
fitting values from the differential ages of passively evolving
galaxies and the dash-dot curves represent $t_{\rm L}$ for best
fitting values from the lookback time to galaxy clusters and the age
of the Universe. The dashed and the dotted curves respectively
correspond to $1\sigma$ and $2\sigma$ errors of $H$ and $t_{\rm L}$
from the differential ages.} \label{fig5}
\end{figure}

We also calculate the corresponding errors $H$ and $t_{\rm L}$
derived from the differential ages of passively
evolving galaxies. The error at different redshift $z$ can be expressed as
\begin{equation}
\sigma^2(z)=\sum^K_{i,j=1}\bigg(\frac{\partial y(z)}{\partial
X_i}\frac{\partial y(z)}{\partial
X_j}\bigg)_{X=\overline{X}}COV(\overline{X_i},\overline{X_j})
,\label{eq16}
\end{equation}
where $K$ is the number of the fitting parameters, $y(z)$ the
theoretical function (such as $H(z)$ and $t_{\rm L}(z)$ here), $X$
and $\overline{X}$ the vectors of fitting parameters and the best
fitting values respectively and $COV$ is the covariance matrix.
We plot in Fig.\ref{fig5} the $1\sigma$ and $2\sigma$ error bars for
$H$ (the left panel) and $t_{\rm L}$ (the right panel) for the
differential ages together with the best fitting results of $H$ and
$t_{\rm L}$ from the lookback time to galaxy clusters and the age of
the Universe. Clearly, the Hubble parameters derived from the two
independent methods are quite consistent with each other. The
corresponding values of $H$ derived from the lookback time to galaxy
clusters and the age of the Universe is nearly within $1\sigma$
region during most of the considered redshift range. For $t_{\rm
L}$, a good agreement between the results from two methods appear at
both lower redshifts at $1\sigma$ level and at higher redshifts at
$2\sigma$ level.

\section{The Deceleration Parameter and The Equation of State}
\begin{figure}
\centering \epsscale{0.9}
\includegraphics[width=15.5cm]{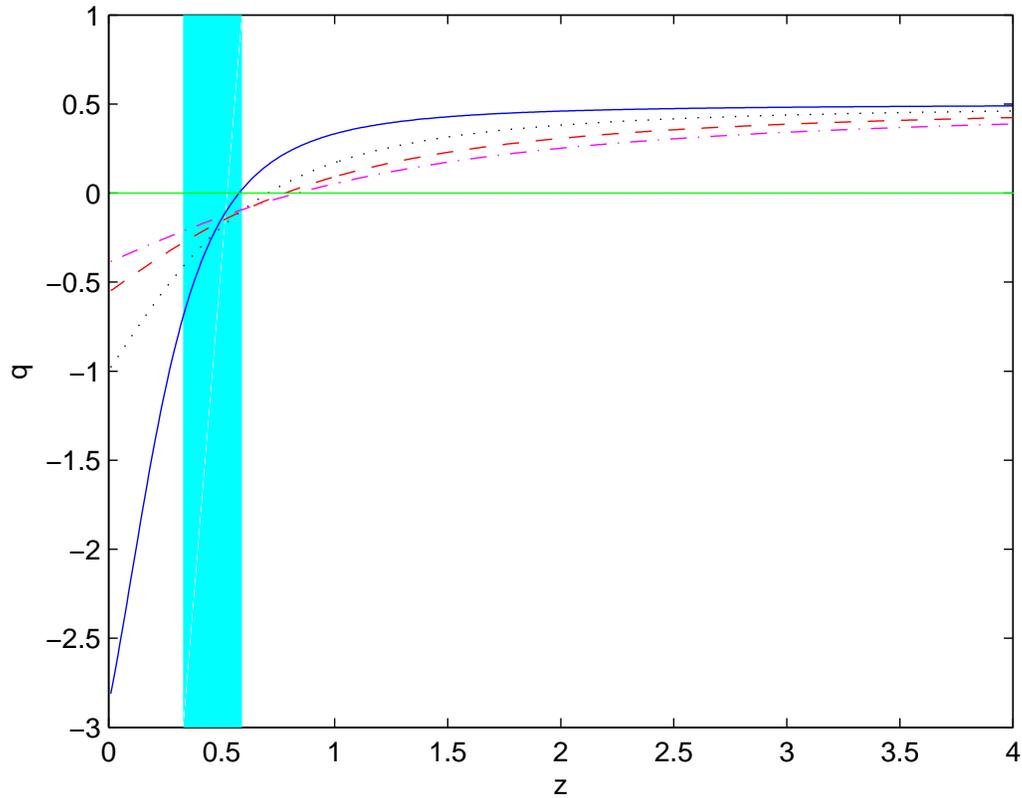}
\caption{Deceleration parameter $q$ as a function of $z$. The dot
line corresponds to $c=0.6$, $\Omega_{\rm m0}=0.24$, the dash
$c=1.0$, $\Omega_{\rm m0}=0.23$ and the dash-dot $c=1.4$,
$\Omega_{\rm m0}=0.21$ respectively. The solid line represents the
case of setting $c$ free, $c=0.26$, $\Omega_{\rm m0}=0.16$ and the
vertical shaded area indicates the SN Ia data constraint on the
transition redshift ($z_{\rm T}=0.46\pm 0.13$) [28].} \label{fig6}
\end{figure}

The deceleration parameter is given in terms of $\Omega_{\rm H}$
like this [15,32],
\begin{equation}
q=\frac{1}{2}-\frac{1}{2}\Omega_{\rm H}-\frac{1}{c}\Omega_{\rm
H}^{3/2},\label{eq17}
\end{equation}
which satisfies $-1/c\leq q\leq 1/2$ due to $0\leq \Omega_{\rm
H}\leq 1$. We plot $q$ as a function of $z$ in Fig.\ref{fig6} using
the best fitting values from the differential ages of passively
evolving galaxies. Obviously $q$ start at the same point $q=1/2$ in
the distant past for all cases of $c$, and the deceleration
parameter is quite sensitive to $c$. Fig.\ref{fig6} also shows that
the values of deceleration parameter today are $q_{\rm 0}$=-2.83,
-0.98, -0.55 and -0.39 for $c$=0.26, 0.6, 1.0 and 1.4 respectively,
which reveal that all the cases support a current accelerating
Universe.
And there exists a transition from deceleration to acceleration at
$z_{\rm T}$=0.58, 0.71, 0.78 and 0.84 corresponding to $c$=0.26,
0.6, 1.0 and 1.4 respectively. The transition redshift $z_{\rm T}$
increases with the increase of $c$.
More recent analysis also favor a lower transition [11,33-36]. In
the work of Huang et al. [13], $z_{\rm T}$=0.28 and 0.72 for
$c$=0.21 and 1.0 respectively. Therefore it seems that a smaller $c$
is more possible once again.

The equation of state is expressed in terms of $\Omega_{\rm H}$ like
this [15,32],
\begin{equation}
\omega=-\frac{1}{3}(1+\frac{2}{c}\sqrt{\Omega_{\rm H}}).\label{eq18}
\end{equation}
It's clear that $\omega$ evolves dynamically and $-(1+2/c)/3\leq
\omega\leq-1/3$ due to $0\leq \Omega_{\rm H}\leq 1$. If $c>1.0$, it
will always be larger than -1 so that the Universe avoids entering
the de Sitter phase and the Big Rip phase [37]. If $c=1.0$, the
holographic scenario will behave like the cosmological constant in
the far future. However, if $c<1.0$, it will probably behave like a
quintom-type dark energy whose equation of state is larger than -1
in the past while it's less than -1 near today.

\begin{figure}
\centering \epsscale{0.9}
\includegraphics[width=15.5cm]{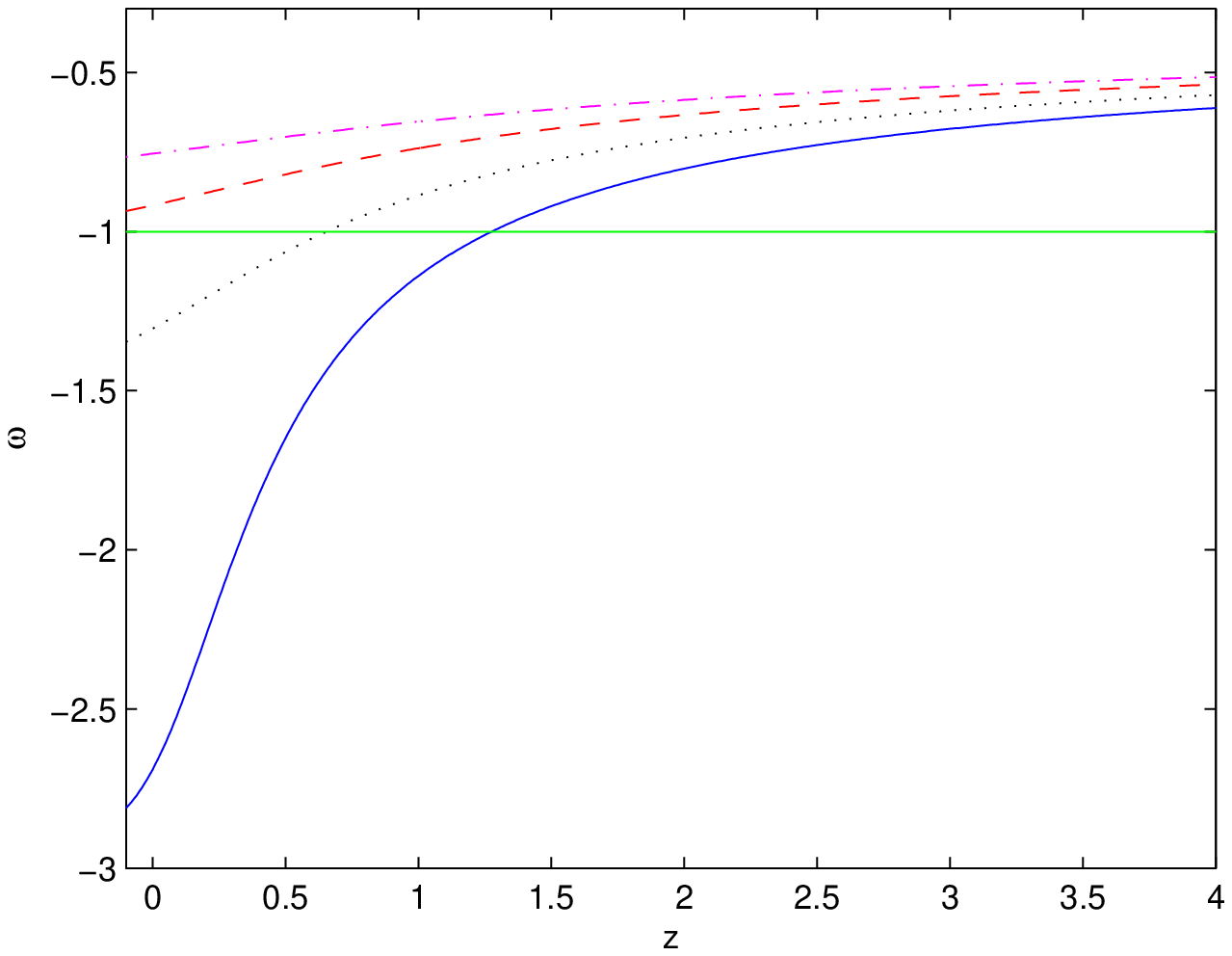}
\caption{Same as Fig.\ref{fig6}, but for the equation of state
$\omega$.} \label{fig7}
\end{figure}

We plot $\omega$ as a function of $z$ in Fig.\ref{fig7} using the
best fitting values from the differential ages of passively evolving
galaxies. We find that at present $\omega _{\rm 0}$=-2.69, -1.30,
-0.92 and -0.75 for $c$=0.26, 0.6, 1.0 and 1.4 respectively.
In the work of Huang et al. [13], $\omega_{\rm 0}$=-2.67 and -0.91
for $c$=0.21 and 1.0, which is slightly larger than
our result of $c$=0.26 and 1.0 respectively. Riess et al. [28]
pointed out $\omega_{\rm 0}<-0.76$ at the 95.4\% confidence level by
using SN Ia data. This result is roughly consistent with our result
of a smaller $c$.

\section{Conclusions and Discussions}
In this paper, we use the the measured Hubble parameters to present
a constraint on a spatially flat Friedmann-Robert-Walker Universe
with a matter component and a holographic dark energy component.
Firstly we consider three fixed values of $c$=0.6, 1.0 and 1.4 in
the fitting of data.
If we set $c$ free, the best fitting values are $c=0.26$, $\Omega
_{\rm m0}=0.16$, $h=0.9998$. It is shown that the holographic dark
energy behaves like a quintom at the $1\sigma$ level although it is
possible for $c>1.0$ at $95.4\%$ confidence level. This result is
consistent with some other independent cosmological tests, which
imply that $c<1.0$ is favored. We also test our results using
the data of lookback time to galaxy clusters and the present age of
the Universe. It is shown that the two independent methods are in
better agreement if the fixed $c$ is smaller and the results from
the two approaches for the case of setting $c$ free are roughly
consistent. We claim that our method is acceptable and our results
are reliable.

Their current values are $q _{\rm 0}=-2.83, -0.98, -0.55, -0.39$
and $\omega _{\rm 0}=-2.69, -1.30, -0.92, -0.75$ for $c$=0.26, 0.6,
1.0 and 1.4 respectively. The transition from deceleration to
acceleration happened at $z_{\rm T}$=0.58, 0.71, 0.78, and 0.84 for
$c$=0.26, 0.6, 1.0, and 1.4.
Compared with the SN Ia observation [28], we find that a smaller $c$
is more acceptable.

It is important to note that this method is much more reliable than
that based on absolute galaxy ages which is more sensitive to
systematics errors [38,39]. Absolute galaxy ages can provide only a
lower limit to the age of the Universe [38].
We are still faced with the
problems of how to select the passively evolving galaxies, how to
confirm the absolute ages and how to determine $\Delta z$. They are
both sticking points but model-dependent, with unavoidable
uncertainties. The data we use were obtained assuming a single-burst
stellar population model and a single metallicity model. In fact,
there may exist a serious warp. While the single metallicity
approximation affects the recovered ages well below the error bars,
the single-burst approximation may not work well generally. The star
formation activity at low redshift for massive elliptical galaxies
could bias the single-burst equivalent ages of the spectrum towards
younger ages.

There still exist some deficiencies waiting for improving for our
work. The amount of data for the differential ages of passively evolving galaxies
 is limited. The best fitting values $\Omega_{\rm m0}=0.16$ and $h=0.9998$ are far from the most known priors
and the confidence regions in the $\Omega_{\rm m0}-c$ plane are not
small enough.
Probably the fitting result can be improved by combining other observational data.
\acknowledgments We are very grateful to the anonymous referee for
many valuable comments that greatly improve the paper. Z. L. Yi
would like to thank Qiang Yuan and Jie Zhou for their valuable
suggestions and Jian Gao for his kindly help. This work was partly
supported by the National Science Foundation of China (Grants
No.10473002 and 10533010), the 985 Project and the Scientific
Research Foundation for the Returned Overseas Chinese Scholars,
State Education Ministry.



\end{document}